\documentclass[11pt,a4paper]{scrartcl}

\usepackage{CLICdp}
\usepackage{amsmath}
\usepackage{tikz}
\usetikzlibrary{shapes.callouts}
\usepackage{tkz-euclide}
\usepackage{CLICdp_definitions}

\usepackage{enumitem}
\setlist[description]{%
  topsep=30pt,               
  itemsep=5pt,               
  font={\mdseries\ttfamily}, 
}

\newcommand{\EPad}{\ensuremath{E_{\mathrm{Pad}}}\xspace}
\newcommand{\ECluster}{\ensuremath{E_{\mathrm{Cluster}}}\xspace}

\newcommand{\rPad}{\ensuremath{R_{\mathrm{Pad}}}\xspace}
\newcommand{\rCluster}{\ensuremath{R_{\mathrm{Cluster}}}\xspace}

\newcommand{\phiPad}{\ensuremath{\phi_{\mathrm{Pad}}}\xspace}
\newcommand{\phiCluster}{\ensuremath{\phi_{\mathrm{Cluster}}}\xspace}

\newcommand{\E}{\ensuremath{\vec{E}}\xspace}
\renewcommand{\pad}{\ensuremath{\mathrm{Pad}}\xspace}

\newcommand{\EevRe}{\ensuremath{\E^{\mathrm{Event}}_{\mathrm{Remaining}}}\xspace}
\newcommand{\EevTot}{\ensuremath{\E^{\mathrm{Event}}_{\mathrm{Total}}}\xspace}

\newcommand{\Eav}{\ensuremath{\vec{E}^{\mathrm{Average}}}\xspace}

\newcommand{\Ecut}{\ensuremath{E^{\mathrm{Cut}}_{\mathrm{Pad}}}\xspace}

\newcommand{\Emin}{\ensuremath{E^{\mathrm{Cut}}_{\min}}\xspace}

\newcommand{\sigmacut}{\ensuremath{N_{\sigma}}\xspace}
\newcommand{\sigmaback}{\ensuremath{\sigma^{\mathrm{Pad}}_{\mathrm{BKG}}}\xspace}

\title{High Energy Electron Reconstruction in the BeamCal}

\clicdpnote{2016}{005}  

\date{\formatdate{25}{10}{2016}}

\addauthor{A.~Sailer}{\institute{1}}
\addauthor{A.~Sapronov}{\institute{2}}

\addinstitute{1}{CERN, Switzerland}
\addinstitute{2}{JINR, Russia}

\onbehalfof{the CLICdp collaboration}

\abstract{%
This note discusses methods of particle reconstruction in the forward region
detectors of
future \mbox{\Pep\Pem} linear colliders such as ILC or CLIC\@. At the nominal luminosity the innermost
electromagnetic calorimeters undergo high particle fluxes from the
beam-induced background. In this prospect, different methods of the background
simulation and signal electron reconstruction are described.
}

\graphicspath{{./logos/}{.}{figures}}

\tikzset{%
  level/.style   = {ultra thick, blue },
  connect/.style = {dashed, red },
  notice/.style  = {draw, rectangle callout, callout relative pointer={#1},
    callout pointer width=1.4 mm, callout pointer shorten=-1.5mm },
  label/.style   = {text width=2cm }
}

\begin{document}

\titlepage{}


\section{Introduction}
\label{sec:intro}


The future TeV energy range \Pep\Pem{} colliders (ILC~\cite{ilctdrVol4},
CLIC~\cite{CLICCDR_vol1}) are expected to
become sensitive probes for potential new physics processes or at least
significantly contribute to the validation of the Standard Model (SM). Many Beyond
Standard Model (BSM) searches
have $t$-channel SM processes as a background~\cite{Fuster:2009em}, the
rejection of this background motivates detector fiducial coverage down to the
smallest possible polar angles.

The BeamCal detector system~\cite{Abramowicz:2010bg} is centred around
the outgoing beam axis in the forward
direction. Its
purposes are: tagging of high energy electrons to suppress backgrounds to potential BSM
process, shielding of the accelerator components from the
beam-induced background, and providing supplementary beam diagnostics
information extracted from the pattern of incoherent-pair energy depositions in
the BeamCal~\cite{Grah:2008zz}.

To achieve nominal luminosities at the level of \SI{d34}{cm^{-2}s^{-1}},
nanometre-sized beams are necessary. The high charge density in the bunches
will induce strong electromagnetic fields causing deflection of the beam
particles and the radiation of \emph{beamstrahlung}. In addition, the
beamstrahlung photons will interact with the beam particles and produce
electron--positron and quark--anti-quark pairs. While the photonic component 
will be radiated practically along the outgoing beam axis, a noticeable
fraction of leptonic and hadronic pairs will hit the BeamCal
calorimeter in the forward region. The distribution of
energy depositions from incoherent pairs depends on the beam parameters and shape and
strength of the detector magnetic field.


Electron tagging at low angles is thus complicated by the high
occupancy in the BeamCal~\cite{sailer2012}. The reconstruction software
for the forward region must include a background-adaptive algorithm in order to
provide maximum tagging efficiency for high energy final state electrons
produced in the collisions.

In this note two such algorithms are presented and
have their performance studied. The first algorithm implements clusterization
of signal energy depositions in the calorimeter shower. The second method is
based on fitting the laterally projected energy distribution with an analytical
formula describing shower energy deposition. The approaches to the background
simulation are also reviewed.

Besides the electron tagging algorithms, the reconstruction software has
several features. It is usable for different detector geometries which can be
defined in the configuration files. It allows tuning of the reconstruction
parameters and presents a choice of several background simulation options. The
code extensibility makes it technically possible for users to implement their
own electron tagging algorithm.


The note is structured as follows: the BeamCal detector design is briefly
described in \Cref{sec:beamcal}, \Cref{sec:beam-bkg} contains a description of
beam-induced background treatment in the simulation and \Cref{sec:reco}
presents two methods of high energy electron reconstruction in the BeamCal, the
algorithm performance and background methods are compared
in \Cref{sec:alg-perf}, \Cref{sec:summary} contains the summary of this study, the
tool for background conversion and the list of simulation options are described
in \Cref{appendix:bkg-files} and \Cref{appendix:config}, respectively.

\section{BeamCal Detector}
\label{sec:beamcal}


The BeamCal is a tungsten-sandwich sampling calorimeter centred on the outgoing
beam-axis. The large dose imparted by the beam-induced backgrounds requires the
use of radiation hard sensors.

\begin{figure}
  \centering
  \begin{subfigure}[b]{0.48\textwidth}
    \includegraphics[width=\textwidth, clip=true, trim=0 5.6cm 13.6cm 2.4cm]{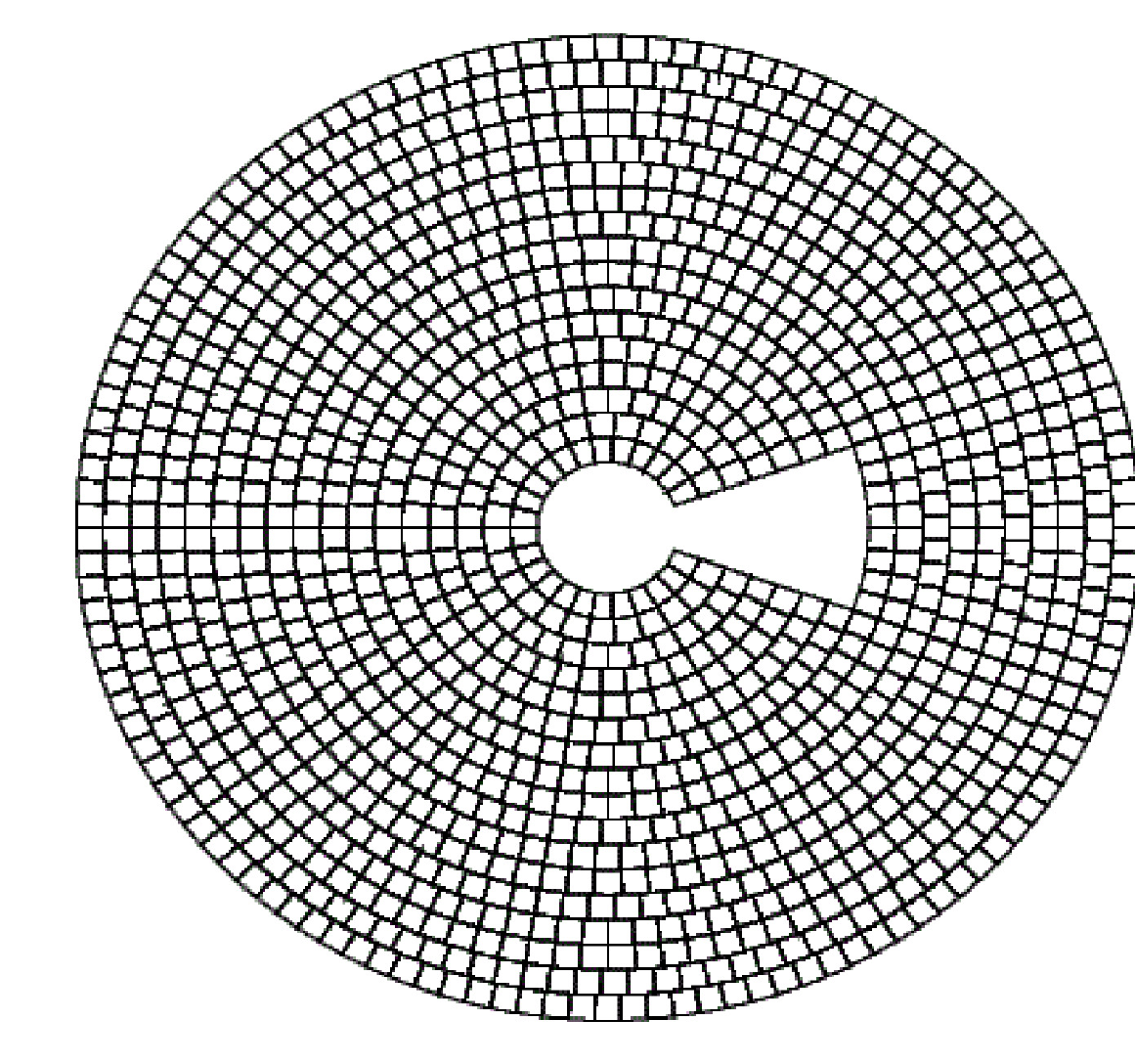}
    \subcaption{}\label{fig:segm-uni}
  \end{subfigure}
  \hfill
  \begin{subfigure}[b]{0.48\textwidth}
    \includegraphics[width=\textwidth, clip=true, trim=14cm 6cm 0 2cm]{./figures/segmentation.pdf}
    \subcaption{}\label{fig:segm-prop}
  \end{subfigure}
  \caption{Different segmentation schemes for the BeamCal calorimeter: uniform
  \subref{fig:segm-uni} and proportional \subref{fig:segm-prop}.}\label{fig:segmentations}
\end{figure}

The choice of the segmentation for the BeamCal sensors influences the
reconstruction efficiency. \Cref{fig:segmentations} shows two different possible
segmentations for the BeamCal: uniform and proportional. In the uniform type of
segmentation the pads have approximately the same size. This type of design is used in
the current studies. In this case the reconstruction efficiency degrades with
lower radii where the background occupancy is higher. An alternative
proportional segmentation without this drawback can be used, but
requires further developments of the simulation.

In the CLIC\_ILD\_CDR~\cite{LCDnote_CLICILDCDRgeo} geometry used for current studies the BeamCal detector has 40 layers of
\SI{3.5}{mm} thick tungsten absorber and radiation-resistant \SI{0.3}{mm} sensor 
uniformly segmented into approximately $8 \times $\SI{8}{mm^2} pads. The R\&D studies to select the
sensor material most suitable for hard radiation environment are
ongoing~\cite{Grah:2009zz,Afanaciev:2012im} and
the current \geant{} simulation uses diamond sensors. \Cref{fig:BeamCal} shows a
render of simulated calorimeter with a \SI{100}{mm} graphite shield to absorb
particles backscattered in the direction of IP\@. The described BeamCal geometry
covers the polar angle span from \SI{10}{mrad} to \SI{43}{mrad}.

The possibility of adjustable geometry and segmentation is accounted for in the
reconstruction framework.

\newcommand{\rotangle}{4.5}
\newcommand{\scale}{17.0}

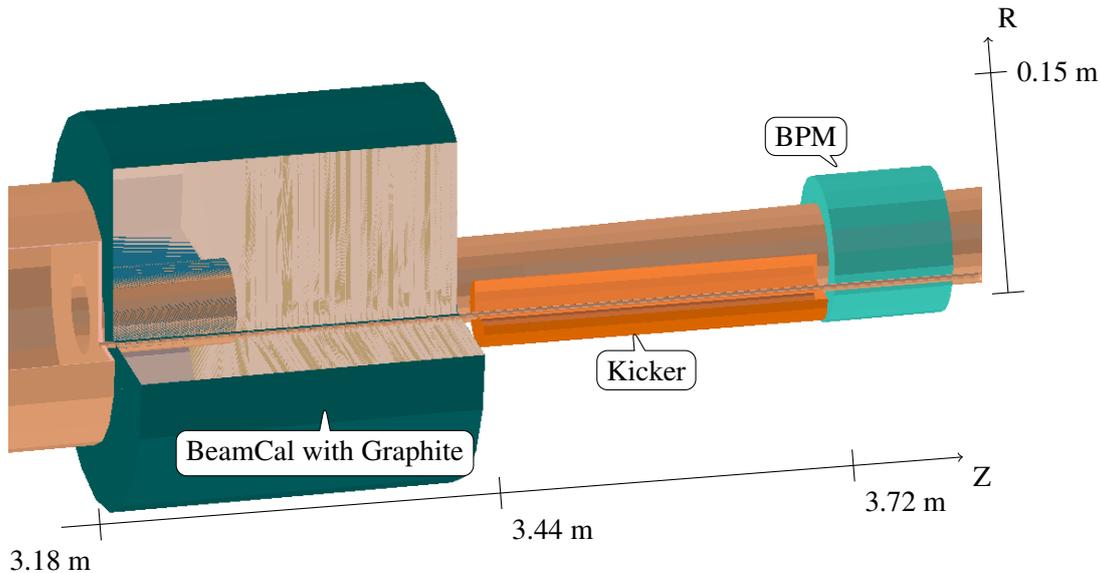
\begin{figure}[bt]
  \centering






  \begin{tikzpicture}

    \node[anchor=south west,inner sep=0] at (0,0) {\includegraphics[width=0.8\textwidth]{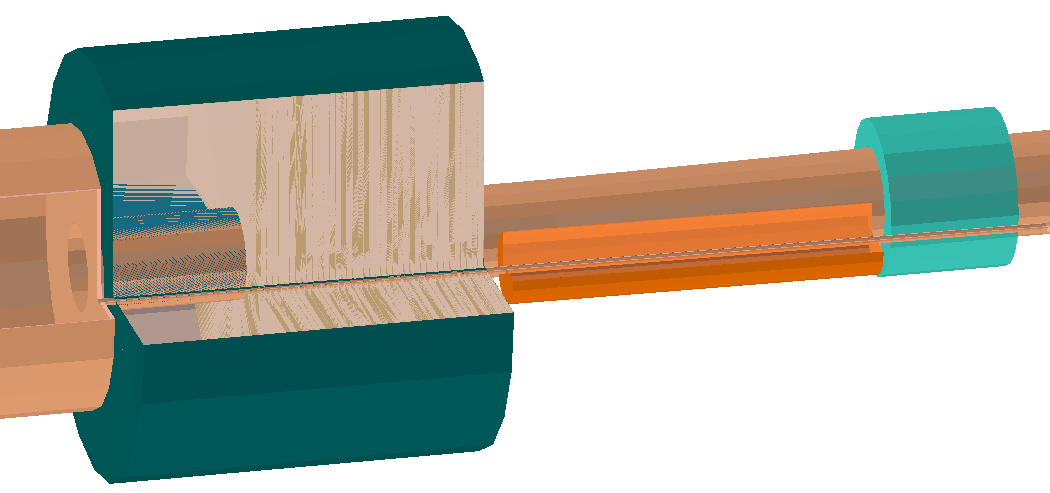}};
    \node[notice={(0.0,0.2)}, rounded corners, fill=white] at (4.17,0.90) {BeamCal with Graphite};
    \node[notice={(-0.05,0.15)}, rounded corners, fill=white] at (8.4,2.00) {Kicker};
    \node[notice={(0.1,-.1)}, rounded corners, fill=white] at (10.5,5.1) {BPM};

    \begin{scope}[cm={cos(-\rotangle) ,-sin(-\rotangle) ,sin(-\rotangle) ,cos(-\rotangle),(-52.0, -2.1)}]

      \draw[scale=\scale,->] (3.10,-0.125) -- ++(0.7,0.0) node[below right] {Z} ;

      \draw[scale=\scale] (3.18-0.05, -0.1125)   -- ++(0.0,-.025) node[below left]  {3.18~m}  ; 
      \draw[scale=\scale] (3.441,-0.1125)   -- ++(0.0,-.025) node[below right] {3.44~m}  ; 
      \draw[scale=\scale] (3.715,-0.1125)   -- ++(0.0,-.025) node[below right] {3.72~m}  ; 

      \draw[scale=\scale,->] (3.845,0.0) -- ++(0.0,0.20) node[above right] {R};
      \draw[scale=\scale] (3.8575,0) node[right]{}--       ++(-0.025,0) ; 
      \draw[scale=\scale] (3.8575,.172) node[right]{0.15~m}-- ++(-0.025,0) ; 

    \end{scope}
  \end{tikzpicture}

  \caption{The BeamCal in a CLIC detector model. Shown is the beam pipe, the
    BeamCal, and the kicker and beam position monitor of the
    intra-train-feedback system. The Z position is given with respect to the
    interaction point, the radial dimension is relative to the outgoing beam
    axis.\label{fig:BeamCal}}
\end{figure}

\section{Simulation of the Beam-induced Background}
\label{sec:beam-bkg}


In the current setup the simulated BeamCal event consists of the energy
depositions from \emph{signal events} and the energy from the background sample.
For simplicity and computational efficiency the signal and background processes
are generated and simulated separately.

To provide the incoherent pair background for the BeamCal reconstruction four
methods were implemented. For each of the background generation methods a set of
simulated background bunch crossings are required as a basis. From this
\emph{background pool} complete bunch crossings are used during the
reconstruction. Alternatively, distributions and parametrisations derived from the background pool can be used.
The procedure to select the background method and set the number of
bunch crossings in the configuration file is described in the \Cref{appendix:config}.

\subsection{Pregenerated Background}
\label{sec:preg-bg}
The first method uses a Monte Carlo technique and is named `pregenerated'. The
event background sample is constructed from the corresponding number of bunch
crossings occurring within the read-out time window randomly selected from the
background pool.
This method gives the most precise and realistic description of the background.
However, the background pool -- consisting of hundreds of bunch
crossings -- 
has to be provided during the reconstruction. This approach is useful to estimate the
BeamCal reconstruction efficiency, but the total file sizes can be prohibitive for
large scale Monte Carlo campaigns.

The procedure to convert a simulated bunch
crossing into the background root file is described in \Cref{appendix:bkg-files}.

\subsection{Parametrised Background}
\begin{figure}
  \centering
  \begin{subfigure}[b]{0.32\textwidth}
    \includegraphics[width=\textwidth]{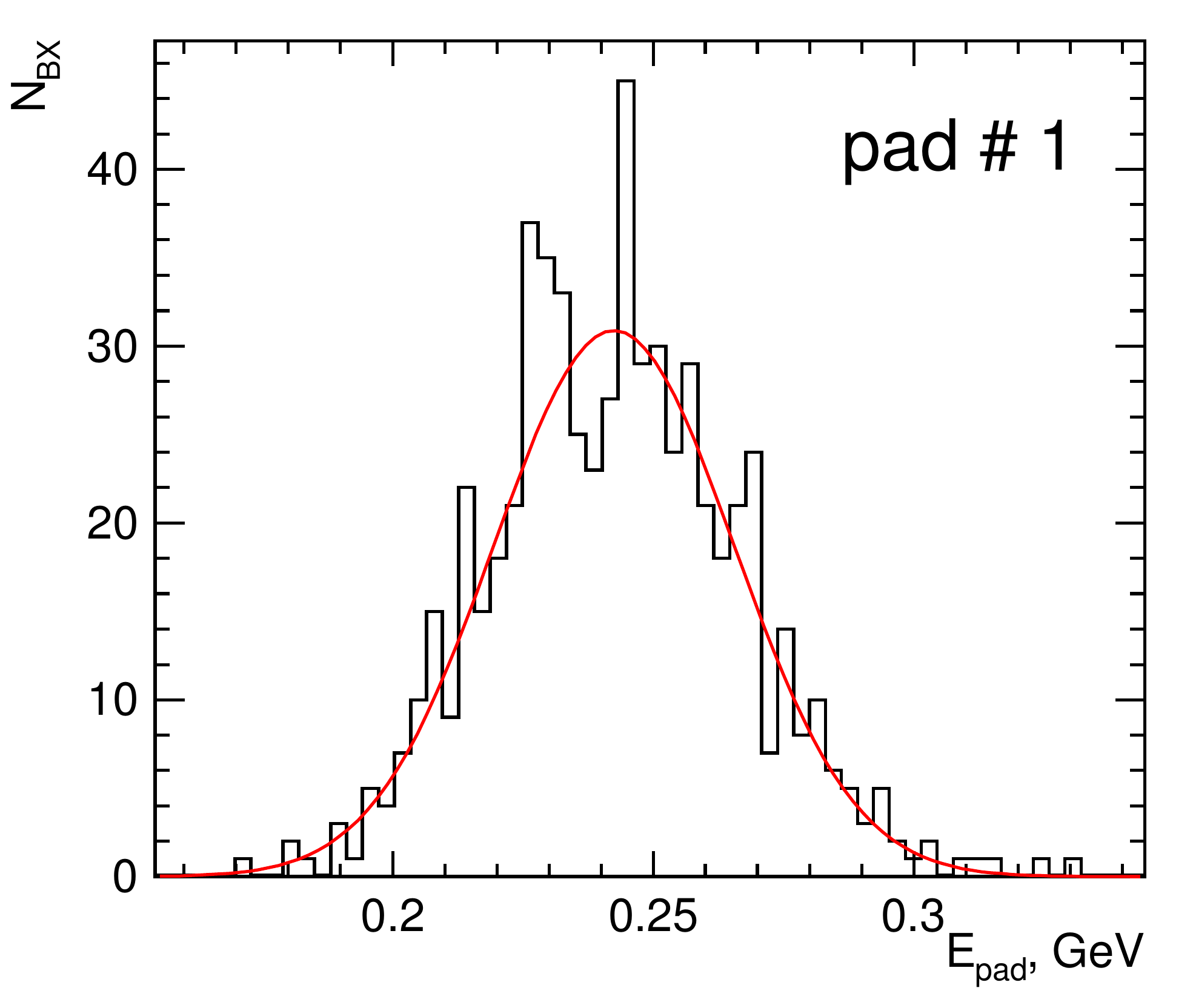}
    \subcaption{}\label{fig:pad1}
  \end{subfigure}
  \begin{subfigure}[b]{0.32\textwidth}
    \includegraphics[width=\textwidth]{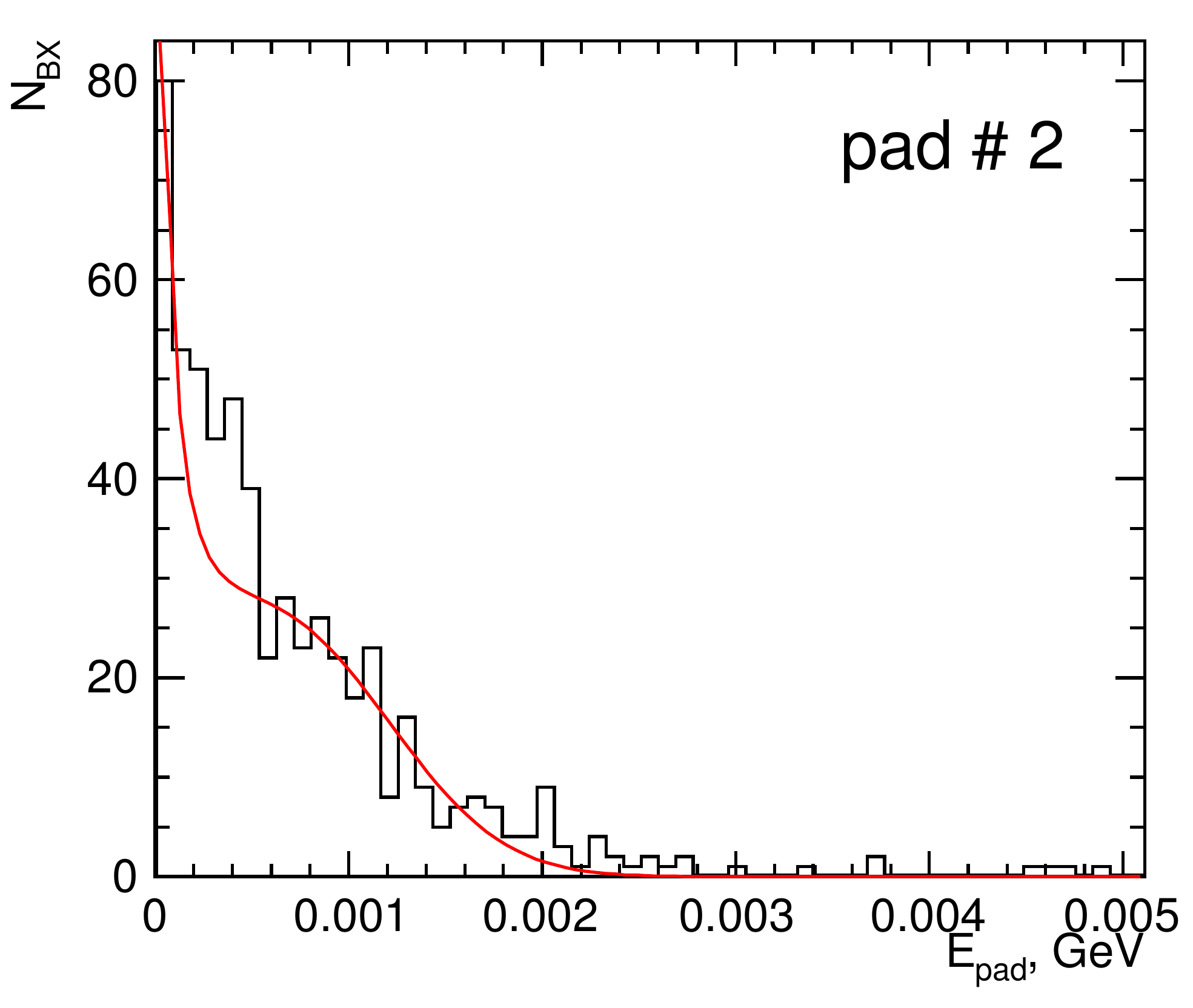}
    \subcaption{}\label{fig:pad2}
  \end{subfigure} \\
  \begin{subfigure}[b]{0.32\textwidth}
    \includegraphics[width=\textwidth]{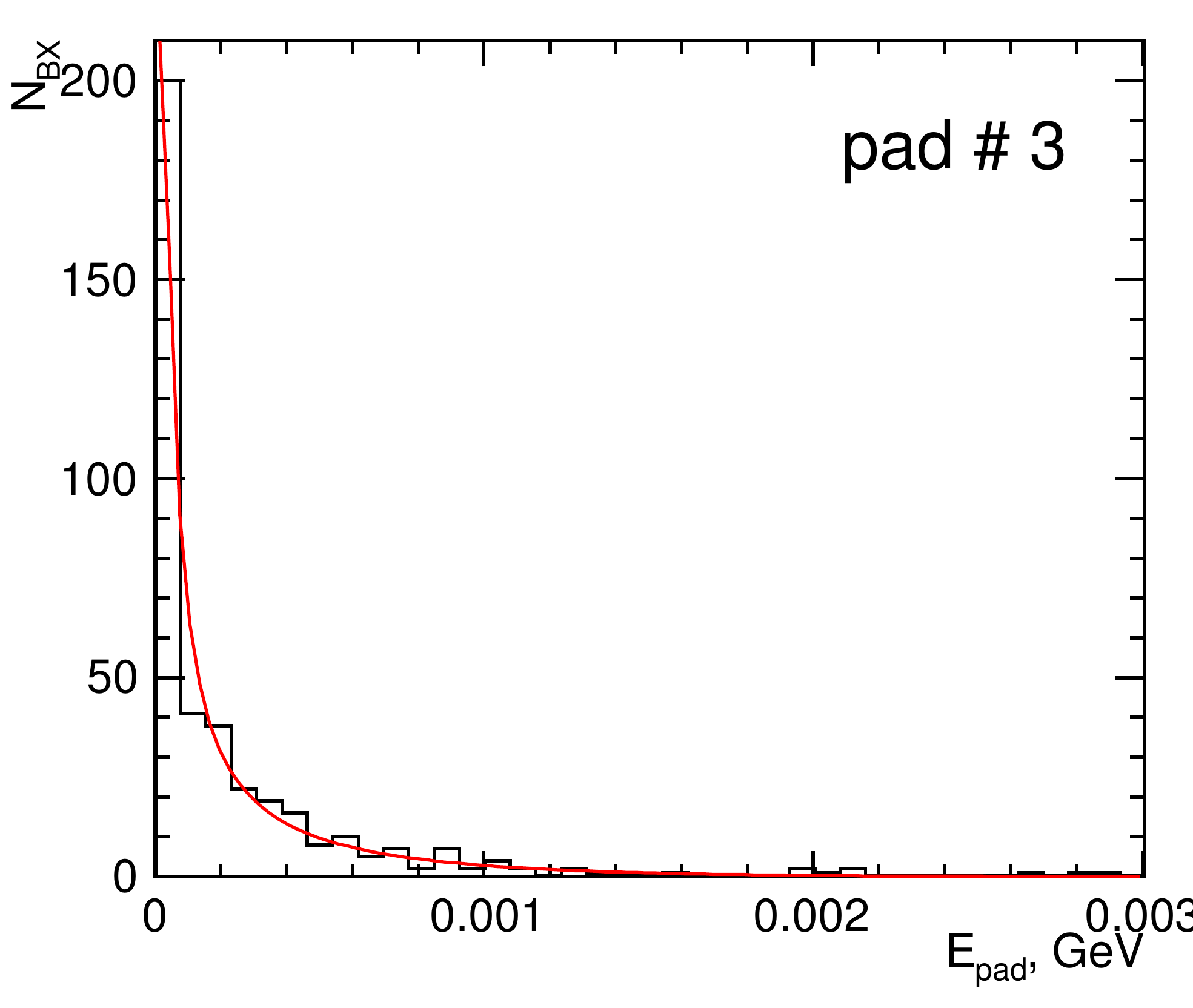}
    \subcaption{}\label{fig:pad3}
  \end{subfigure}
  \begin{subfigure}[b]{0.32\textwidth}
    \hspace{0.5cm}
    \begin{tikzpicture}
    \node[anchor=south west,inner sep=0] at (0,0) {\includegraphics[width=0.85\textwidth, height=0.73\textwidth, clip=true, trim=7cm 3.0cm 3.4cm 5cm]{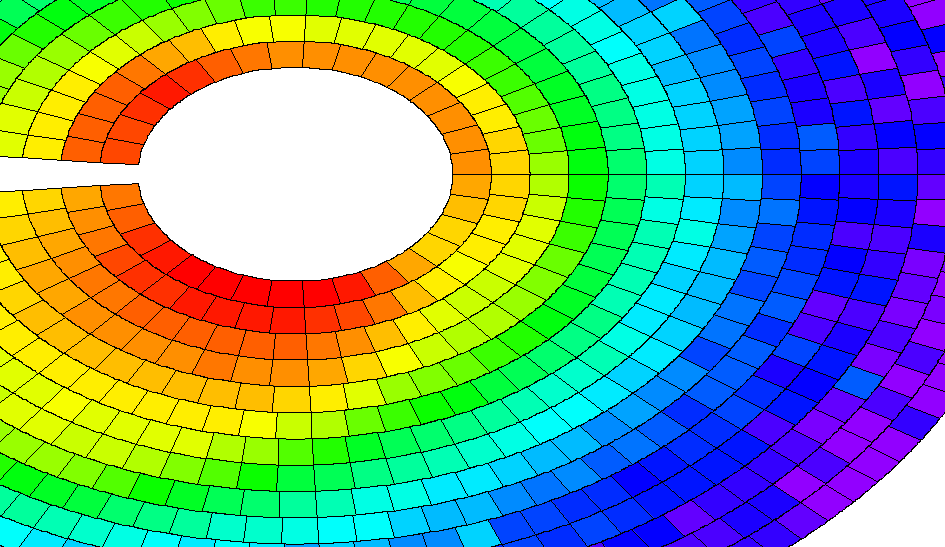}};
    \node[notice={(0.0,0.2)}, rounded corners, fill=white] at (1.17,1.10) {\scriptsize \#1};
    \node[notice={(-0.05,0.15)}, rounded corners, fill=white] at (2.38,1.30) {\scriptsize \#2};
    \node[notice={(0.0,-0.2)}, rounded corners, fill=white] at (3.37,1.75) {\scriptsize \#3};
    \end{tikzpicture}
    \vspace{0.48cm}
    \subcaption{}\label{fig:allpads}
  \end{subfigure}
  \caption{\subref{fig:pad1}--\subref{fig:pad3} The distribution of the
    background energy deposition for bunch crossings at \SI{3}{\TeV} CLIC in three distinct pads fitted with
    the parametrisation. \subref{fig:allpads} The corresponding pads marked at the BeamCal
    front projection of layer 10. The plots show that
    different background energy spectra are well described by the same
    parametrisation.  }\label{fig:param-bg}
\end{figure}
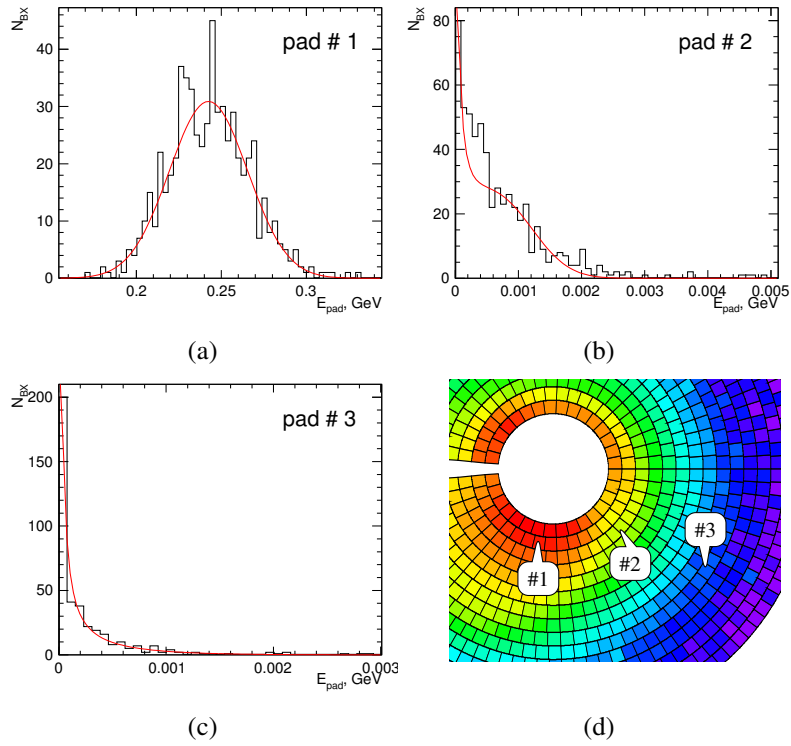

 An alternative method of providing the energy deposition in the BeamCal by background particles is called
`parametrised', where the background energy deposition in each pad is generated
according to the distribution
\begin{equation*}
F(x) = \frac{A}{x}\exp{\left(\frac{x-B}{C}\right)}^2
\end{equation*}
where the $A, B$ and $C$ parameters are determined for each pad by fitting
the energy depositions from the background pool.
An example of background energy distributions 
for three selected pads is shown with the corresponding fitted functions on
\cref{fig:param-bg}. The pads were selected to represent three distinct energy
spectra: pad \#1 in \cref{fig:pad1} with a quasi-symmetric Gaussian, pad \#2 in \cref{fig:pad2}
with a Gaussian near 0 and pad \#3 in \cref{fig:pad3} with lower energy
depositions proportional to $1/x$. The pad
positions are marked on the front view of the first sensor plane in \cref{fig:allpads}. All
three sample spectra are reasonably well described by the provided
parametrization.

\begin{figure}
  \centering
  \begin{subfigure}[b]{0.48\textwidth}
    \includegraphics[width=\textwidth]{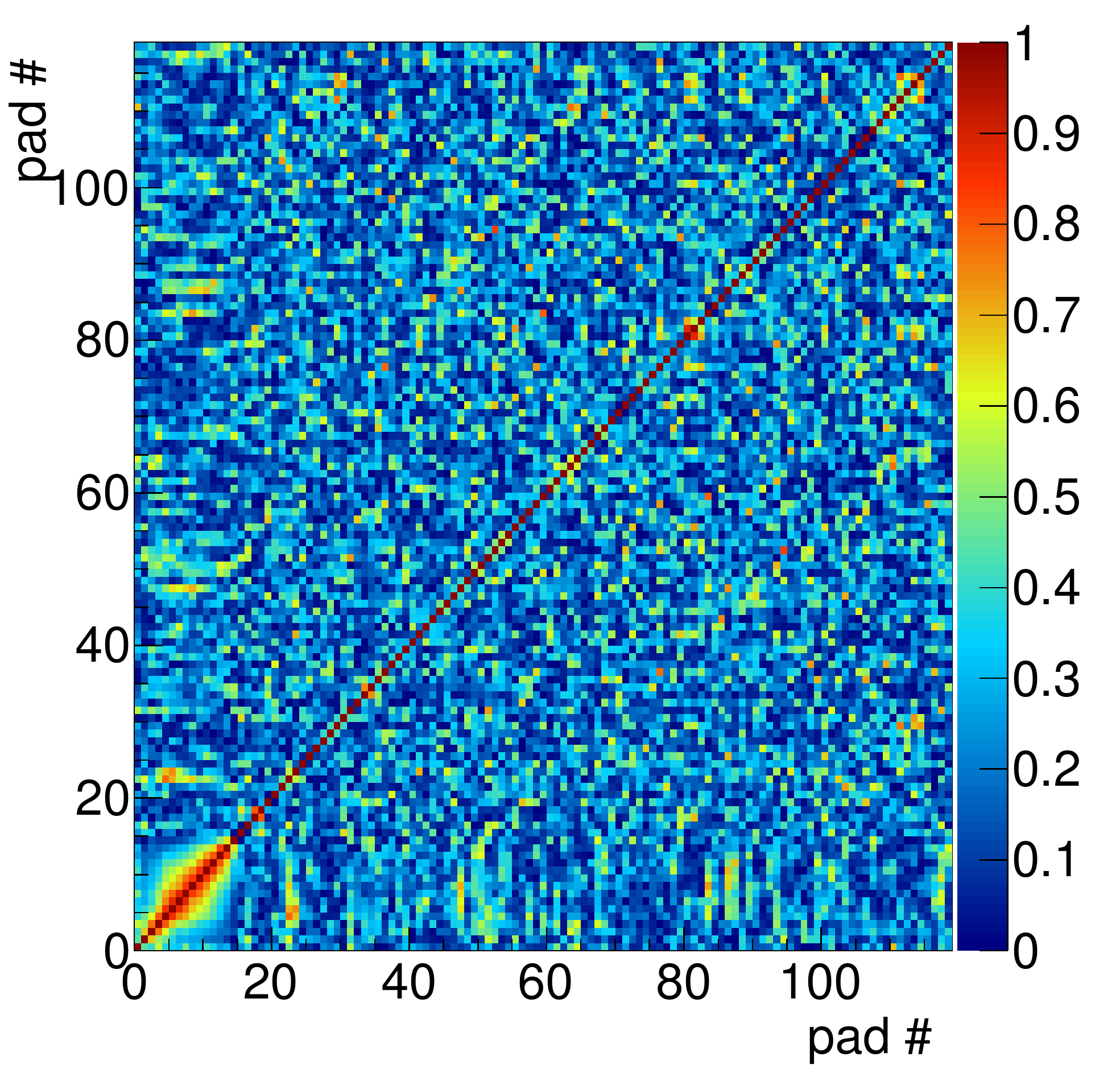}
    \subcaption{}\label{fig:layer-corr}
  \end{subfigure}
  \hfill
  \begin{subfigure}[b]{0.48\textwidth}
    \includegraphics[width=\textwidth]{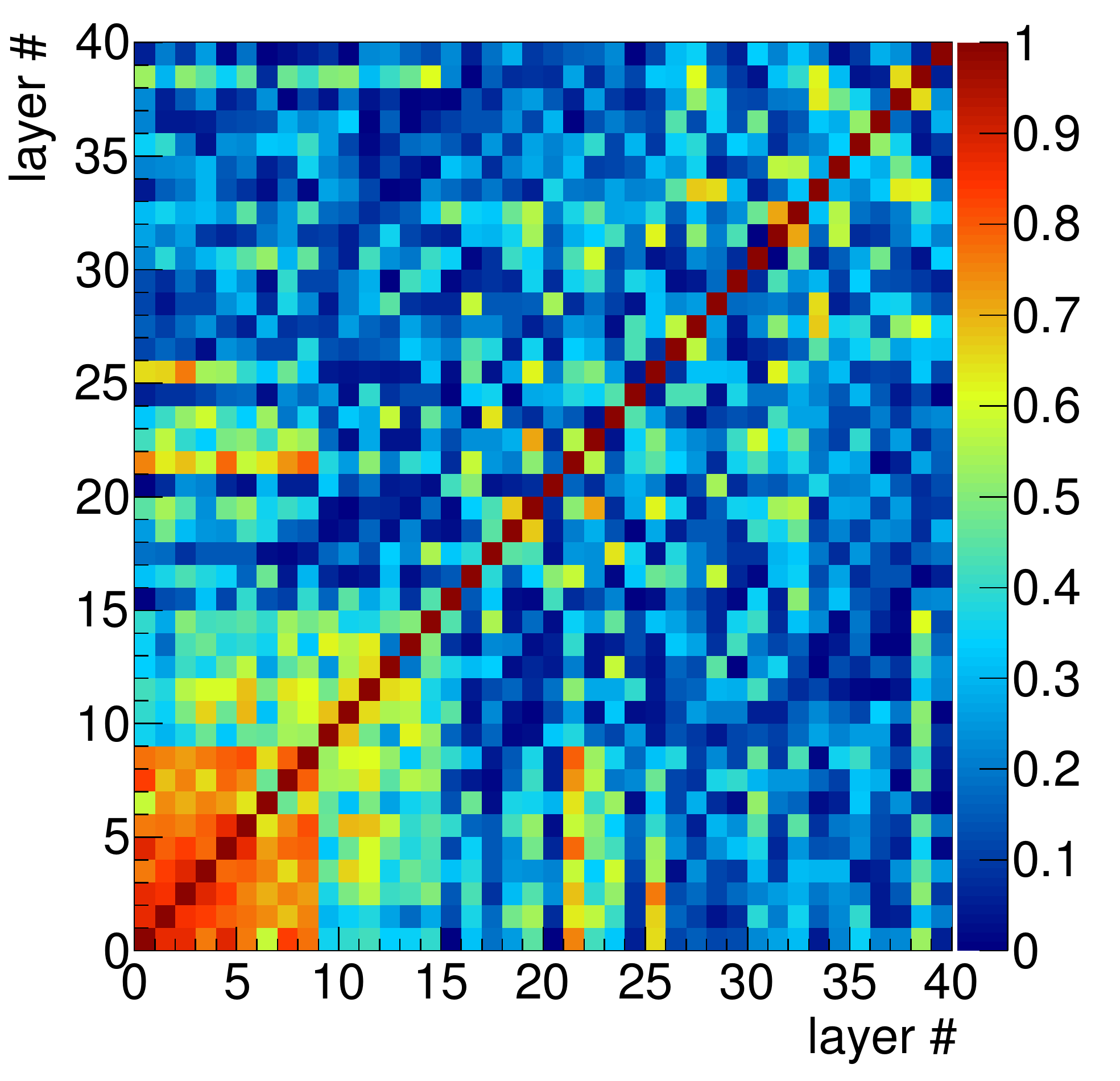}
    \subcaption{}\label{fig:tower-corr}
  \end{subfigure}
  \caption{Background energy deposition correlations between pads in three
  inner rings of the $10^{\mathrm{th}}$ layer \subref{fig:layer-corr} and along the BeamCal axis
  for one of the innermost pads \subref{fig:tower-corr}. The corresponding pad or layer
  numbers are given along the axes. The backgrounds are for CLIC at \SI{3}{\TeV}.\label{fig:bg-corr}}
\end{figure}

The parametrised method gives a result that is almost as precise as the `pregenerated' method
except for the correlations between the energy deposits in neighbouring pads.
The plot in \cref{fig:layer-corr} shows the correlation matrix between the
120 innermost pads in the front projection representing the area with the highest energy
deposition (the first three rings). The correlation matrix on \cref{fig:tower-corr}
is drawn for one of the innermost towers of pads along the detector axis. The plots
show that the correlations are small in most parts of the detector and
especially in the region normally used for the reconstruction (after the tenth
layer). The correlations can therefore be neglected for the parametrised and the Gaussian
methods of the background generation. 

Due to the large number of pads in the BeamCal the generation time becomes very
long for a large number of bunch crossings. This method is thus applicable for
cases where only a few bunch crossings take place within the read-out time window.

The procedure to convert a simulated bunch
crossing into the background file usable with this and the Gaussian approach is
described in \Cref{appendix:bkg-files}.

\subsection{Gaussian Background}
\label{sec:gauss-bg}
 Another method of producing background is called
`Gaussian', where the background energy deposition in each pad is generated
according to a Gaussian distribution with a mean of $\bar{E}_{\pad}$ and a variance
$\sigma_{\pad}^2$ determined from the
background pool. Although for single bunch crossings the distribution of energy
deposition in a pad differs from a Gaussian distribution, for sufficiently large number of bunch crossings $N > 10$
the distribution of their sum will be well described by a Gaussian with the mean
$N\bar{E}_{\pad}$ and the variance $N\sigma_{\pad}^2$, according to the Central
Limit Theorem. This method is thus
applicable for the read-out samples over a large number of bunch crossings.

\subsection{Averaged Background}
The fourth possibility is similar to using the Gaussian background. The method
is provided in the simulation for backward compatibility with the electron
tagging reconstruction in \textsc{MarlinReco}~\cite{marlinreco} and can read the files with
averaged energy density used in that processor. The background distribution is generated from the averages.

\section{High Energy Electron Reconstruction}
\label{sec:reco}

The techniques of shower position reconstruction in laterally segmented
calorimeters were developed and presented
in~\cite{Akopdzhanov:1976pr, Bugge:1986mq,Awes:1992yp}. They essentially converge
to two methods: a clustering algorithm based on calculation of the centre of gravity of signal
pads, previously used in the FCal collaboration~\cite{fcal2004};
and a method based on fitting the energy deposition with a modelled shower shape. While
the first
method is simpler and faster, it is optimised for perpendicularly segmented
sensors. In case of the radial segmentation, the fitting method may have better
performance in terms of precision. However the main purpose of the BeamCal
detector is the tagging of high-energy electrons, while measuring their spatial
coordinates has lower priority. Therefore the choice between these methods
will have to be made depending on the specific application or analysis.


To perform electron tagging with the BeamCal a reconstruction procedure was
developed. It relies on the two aforementioned algorithms implemented as a \marlin~\cite{MarlinLCCD}
processor \texttt{BeamCalClusterReco} and built into the global detector reconstruction
framework. The processor also takes care of background generation when the
reconstruction is applied to simulated signal.

\subsection{Clustering Algorithm}
\label{sec:clust}

This clustering option is a nearest neighbour search based on the pads with
significant remaining energy after the subtraction of the average background.

\subsubsection{Energy Subtraction}

In the first step of the algorithm the average expected energy from incoherent
pair background \Eav{} is removed from the total energy \EevTot, which is the sum of the signal
and background energy deposits for each pad. Given the remaining energy in
each pad,
\begin{equation*}
  \EevRe = \EevTot - \Eav ,
\end{equation*}
pads for clusters are selected.

\subsubsection{Pad Selection}

As the next step, pads with a significant amount of remaining energy are chosen
for the clustering. There are two options to select pads. A pad selection based
on a constant minimal required energy depending on the ring of a pad, and a pad
selection based on the standard deviation of the background energy deposit in
each pad.

\paragraph{Constant Energy Selection}

A pad is selected for further clustering if the pad energy $\EPad(\mathrm{Ring}_{\pad})$ is larger than the
minimal required energy in its ring $\Ecut(\mathrm{Ring}_{\pad})$
\begin{equation*}
  \EPad(\mathrm{Ring}_{\pad}) > \Ecut(\mathrm{Ring}_{\pad})\,.
\end{equation*}

The pad selection in this case is steered by the \texttt{ETPad} parameter of the
BeamCalReco processor.

\paragraph{Variable Energy Selection}

In this case, pads are selected based on the energy fluctuation of the background
from event to event. The standard deviation of the energy fluctuations for the
background \sigmaback{} is calculated. Pads are selected if the remaining
energy \EPad{} is larger than \sigmacut{} standard
deviations. It is also possible to define a minimal remaining energy \Emin{} to
select only pads which have at least \Emin{} remaining energy
\begin{equation*}
  \EPad > \max( \Emin, \, \sigmacut \cdot \sigmaback )\,.
\end{equation*}

In the processor parameters, \Emin{} is equal to the first value of the
\texttt{ETPad} parameter and \sigmacut{} is given by the \texttt{SigmaCut} parameter.

\subsubsection{Tower Creation and Nearest Neighbour Search}

From the selected pads \emph{towers} are created. A tower is simply
the collection of pads with the same r and $\phi$ coordinates in the
BeamCal. The tower with the \emph{largest} number of selected pads is chosen. The pads in a tower do not have to be in
consecutive layers.  If there are towers next to the primary towers, these are
added to the primary tower, and the added towers are also checked for
neighbours.  Finally, a cluster is created from the tower if there are more than
\texttt{MinimumTowerSize} pads in the cluster. If there are towers not included
in this first cluster, additional clusters might be created until no more towers remain.

\subsubsection{Cluster Location Calculation}

The tower locations are calculated based on the energy-weighted position of the pads in each tower.

The polar angle $\theta_{\mathrm{Reco}}$ is calculated from the average ring
$\rCluster$ of the cluster, where \rPad{} is the radius of a pad
\begin{align*}
  \rCluster = & \frac{1}{\ECluster} \sum_{\mathrm{Pads}} \EPad\, \rPad \, ;\\
\end{align*}
and the azimuthal angle
\begin{align*}
  \phiCluster = & \mathrm{ATan2}\left(
    \frac{1}{\ECluster} \sum_{\mathrm{Pads}} \EPad \sin{\phiPad} ,\qquad
    \frac{1}{\ECluster} \sum_{\mathrm{Pads}} \EPad \cos{\phiPad}\right),
\end{align*}
are calculated
from the energy weighted azimuthal angle \phiPad{} of the pads in the cluster,
where the sums are the average position in $Y$ and $X$, and $\mathrm{ATan2}$ is the
two-argument arc-tangent function commonly found in mathematical libraries.

\subsection{Shower Fitting Algorithm}
\label{sec:shower-fit}

This approach is based on approximating the profile of the high energy electron
shower with a two-dimensional exponential
distribution.  The algorithm utilises a $\chi^2$-test to detect an excess in
the energy deposition over the background. In the present case, i.e., in
simulation, the average
background value and its variance are extracted from the background pool. The
event sample with its total energy deposition is constructed as described in
\Cref{sec:beam-bkg}. The reconstruction is performed within a subset of
calorimeter layers which is defined with a starting layer and depth parameters
in the configuration file.

\subsubsection{Step 1}
The energy depositions in the layers are projected along the calorimeter axis
to its front plane. Each pad in the resulting front projection contains the
following quantities, where the summation is performed over pads in the layers:
\begin{itemize}
\item sum of the total energy depositions behind the pad,
  \begin{align*}
    E_{\proj}^{\tot} =
    \sum E_{\pad}^{\tot} \,;
  \end{align*}
\item sum of the average background depositions,
  \begin{align*}
    \bar{E}_{\proj}^{\bkg} = \sum \bar{E}_{\pad}^{\bkg} \, ;
  \end{align*}
\item sum of the background variance,
  \begin{align*}
    \sigma^2_{\proj} = \sum \sigma^2_{\pad} \,;
  \end{align*}
\item quadratic norm,
  \begin{align*}
    \chi^2_{\proj} = \sum \frac{{(E_{\pad}^{\tot} - \bar{E}_{\pad}^{\bkg})}^{2}}{\sigma^2_{\pad}} \,.
  \end{align*}
\end{itemize}

\subsubsection{Step 2}
After the values are calculated, the algorithm tries to form a shower spot.
It selects a pad from the projection satisfying the following criteria:
\begin{itemize}
\item the pad has the highest $\chi^2_{\proj}$ and it is above a configured threshold
  $\chi^2_{\mathrm{\proj, \min}}$;
\item the difference $E_{\proj}^{\tot} - \bar{E}_{\proj}^{\bkg}$ is above 70\% of the configured energy threshold
  for the total shower energy $E_{\minm}$ defined by the \texttt{ETCluster} parameter and
  described in \Cref{appendix:config}.
\end{itemize}
The selected pad is declared the central pad of the shower. Other pads within $2\rho_{M}$
(Moliere radius) are inspected for the following criteria:
\begin{itemize}
\item the difference $E_{\proj}^{\tot} -
\bar{E}_{\proj}^{\bkg}$ is above 10\% of the configured energy threshold
for the total shower energy $E_{\minm}$ defined by the \texttt{ETCluster}
parameter (see~\Cref{appendix:config});
\item the signal energy measured in standard deviations of the
background $(E_{\proj}^{\tot} - \bar{E}_{\proj}^{\bkg})/\sigma_{\proj}$ is $>1$.
\end{itemize}
Such pads are added as peripheral pads to the spot.

\subsubsection{Step 3}
For simplicity, the shower transverse energy distribution is approximated with
\begin{align*}
E(r) = E_0\exp\left(-\frac{r}{R_0}\right)\,,
\end{align*}
where $r$ is the distance from the shower centre, $E_0$ is a scaling factor and
$R_0$ is the shower width. The shower width varies with
its depth so that $R_0$ depends on the layers from which the energy projection is
calculated. 

The fit is performed with four parameters: the shower centre coordinates $R$ and
$\phi$ in the polar coordinate system with the origin placed at the BeamCal centre,
the scaling coefficient $E_0$, and the shower width $R_0$. Initial values of the centre
coordinates are estimated with the centre-of-gravity method with logarithmic
weights~\cite{Awes:1992yp}. Using numeric integration the algorithm calculates approximate energy
deposition in each spot
pad $E_{\mathrm{int}}$ and calculates a $\chi^2$ measure:
\begin{align*}
\chi^2_{\mathrm{spot}} =
\sum_{\mathrm{spot\;pads}} \frac{{(E^{\mathrm{int}} - (E_{\proj}^{\tot} -
\bar{E}_{\proj}^{\bkg}))}^2}{\sigma^2_{\proj}} \,,
\end{align*}

MINUIT~\cite{James1975} is then used to minimise this value by varying the $R, \phi,
E_0$ and $R_0$ parameters. The resulting $R$
and $\phi$ values are treated as the shower centre and the $E^{\mathrm{int}}$ value
corresponding to the minimum $\chi^2_{\mathrm{spot}}$ as the shower energy.

\subsubsection{Step 4}
Steps 2 and 3 are repeated with the next shower candidate until no more
candidates are found in the front projection.

\section{Algorithm Performance}
\label{sec:alg-perf}

The two algorithms presented in the previous section were compared in terms of
efficiency, fake rate, and spatial and energy resolution. The tests were
performed with beam-induced background simulated for \SI{3}{\TeV} collisions and
signal samples with electron energies from \SI{500}{GeV} to \SI{1500}{\GeV}.
The beam
energy of \SI{1.5}{\TeV} corresponds to the highest and most challenging background
occupancy in the BeamCal. Two methods of background generation were tested:
\emph{pregenerated} and \emph{Gaussian}. In the reconstruction the read-out
window was set to \SI{40}{BX} which means that every signal event was overlaid
with background energy deposition accumulated during 40 bunch crossings.

To test the selection efficiency, mono-energetic electrons and incoherent pair background
were simulated separately in the \geant-based \mokka\footnote{The reconstruction
has since been adapted to work with geometry and simulation based on the
DD4hep geometry framework.}
framework~\cite{Mora2002}.
For the incoherent pair background at the \SI{3}{\TeV} CLIC, beam-beam
interactions with irregular beam shapes were simulated with the \guineapig{} Monte
Carlo program~\cite{schulte1996}. Each element of the background collection
corresponds to a single bunch crossing (BX) with different random seeds and
different initial beam-particle distributions.

To produce an event for the analysis sample the signal energy deposition
($E^{\mathrm{signal}}$) was overlaid on top of the incoherent pair background ($E^\bkg$)
which would accumulate within the read-out window.

\begin{figure}
  \centering
  \begin{subfigure}[b]{0.48\textwidth}
    \includegraphics[width=\textwidth]{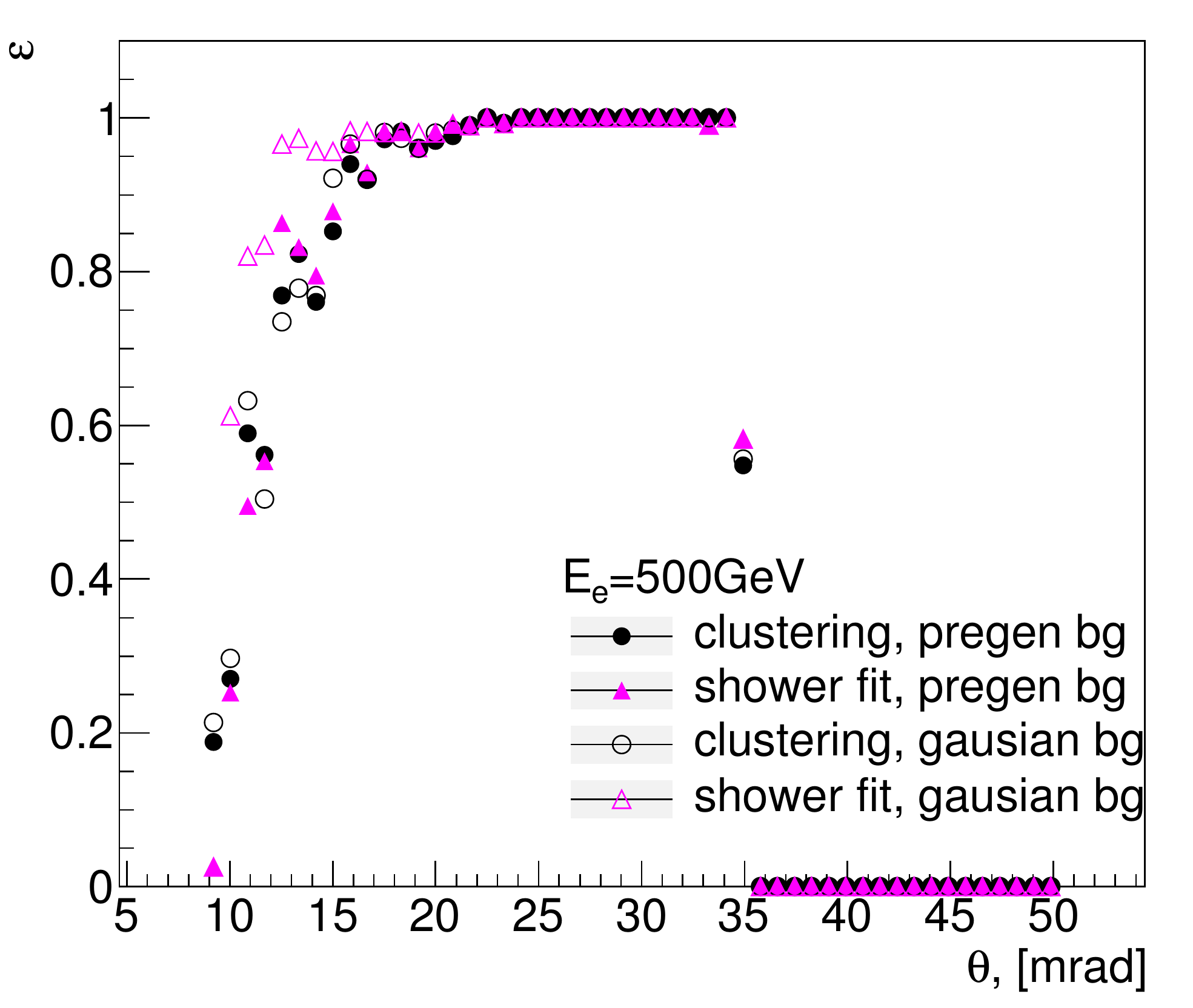}
    \subcaption{}\label{fig:theta-eff-500}
  \end{subfigure}
  \hfill
  \begin{subfigure}[b]{0.48\textwidth}
    \includegraphics[width=\textwidth]{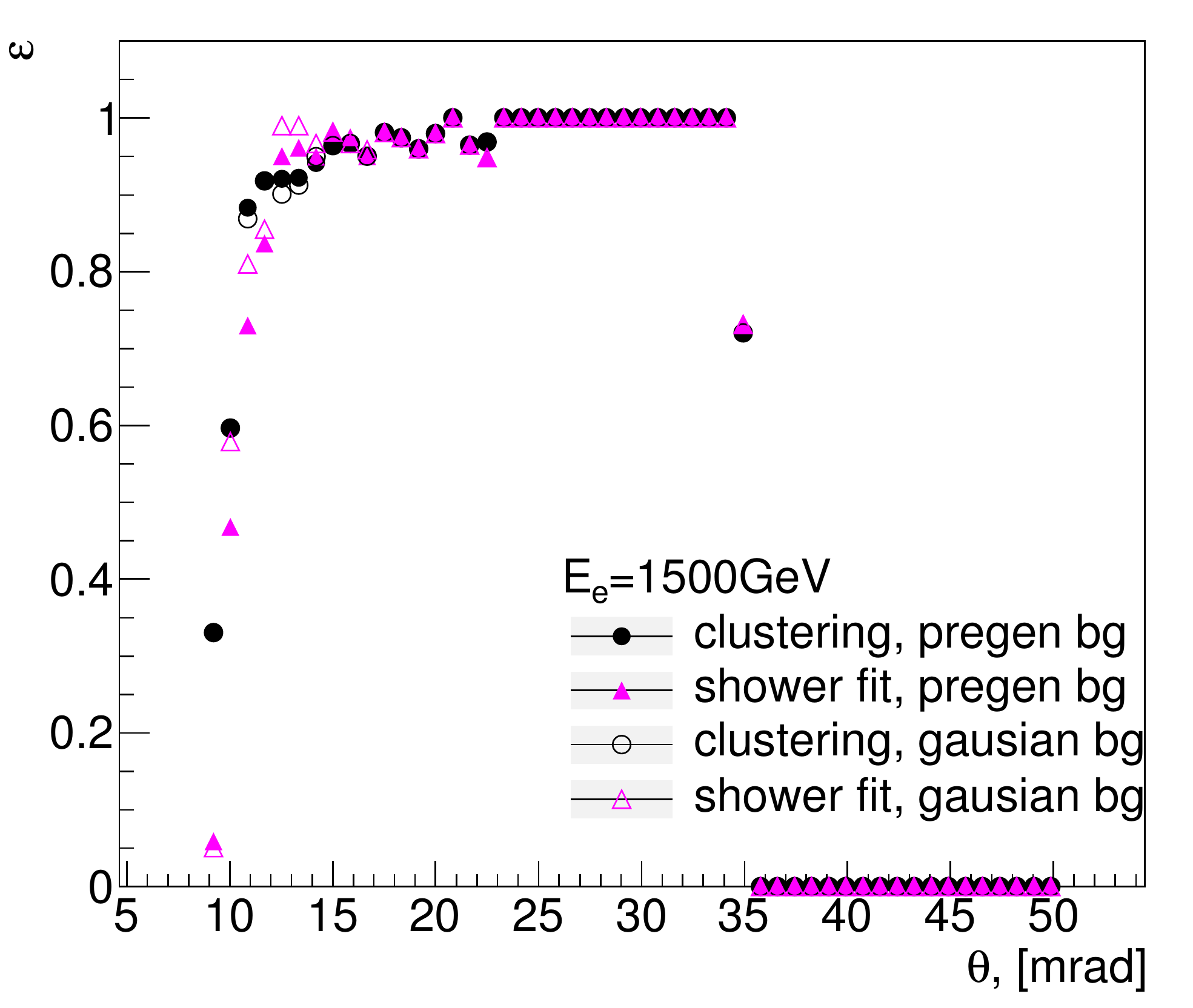}
    \subcaption{}\label{fig:theta-eff-1500}
  \end{subfigure}
  \caption{Polar angle dependence of the reconstruction efficiency for different
  methods. Efficiency for \subref{fig:theta-eff-500} \SI{500}{\GeV} and \subref{fig:theta-eff-1500} \SI{1500}{\GeV} electrons
  hitting BeamCal.\label{fig:theta-eff}}
\end{figure}

The algorithm performance is demonstrated by efficiency plots shown in
\Cref{fig:theta-eff} for different combinations of reconstruction (clustering
and shower fitting) and background
simulation methods (pregenerated and Gaussian). The fraction of detected electrons depends on their
energy and polar angle with respect to the detector axis. At lower $\theta$,
the background occupancy is high and therefore the efficiency is lower.

\begin{figure}
  \centering
  \begin{subfigure}[b]{0.48\textwidth}
    \includegraphics[width=\textwidth]{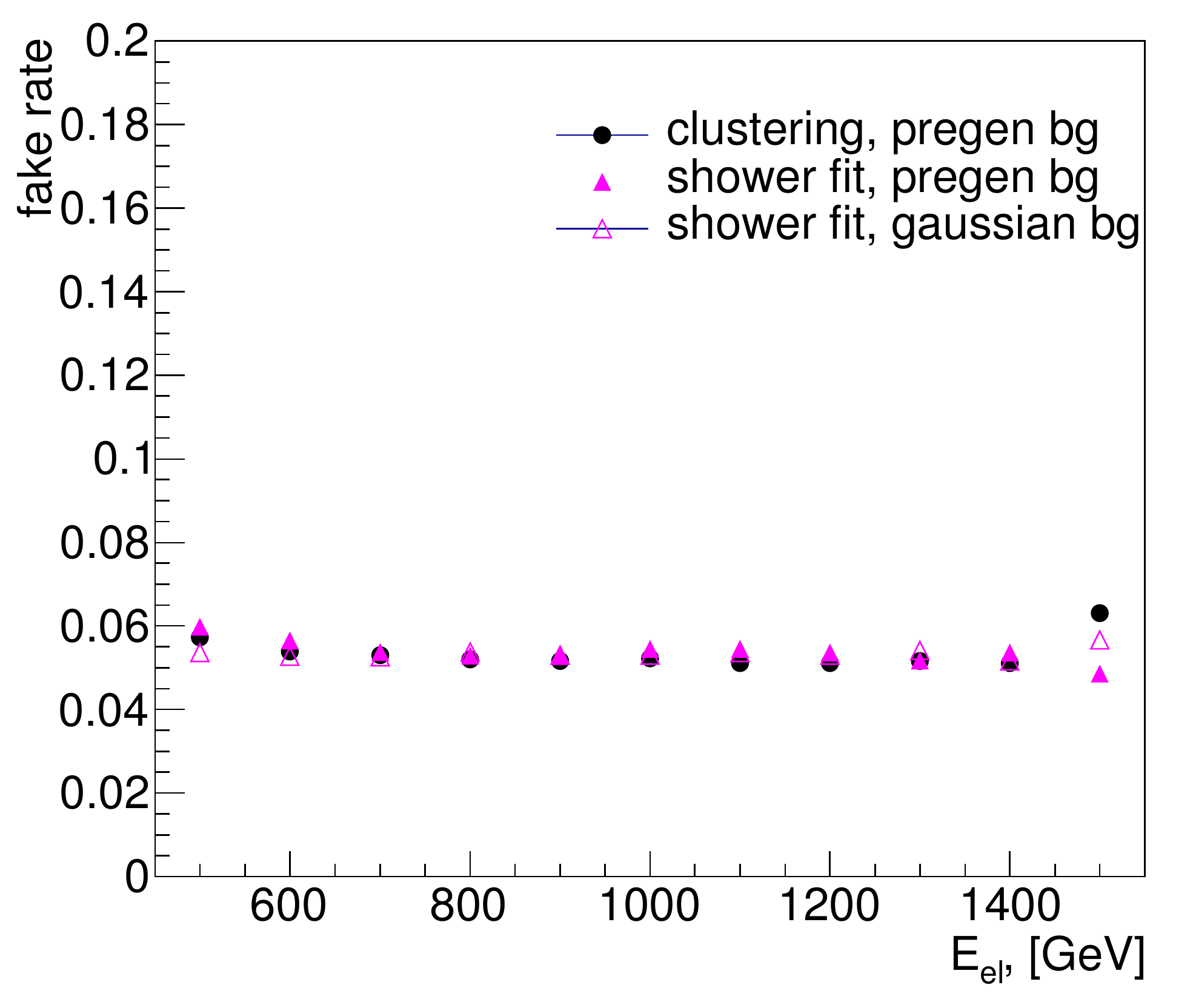}
  \end{subfigure}
  \caption{Dependence of the reconstruction fake rate on the incident electron
  energy.}\label{fig:fake-rate}
\end{figure}

The quality of the reconstruction was compared for three cases: clustering
algorithm with pregenerated background and shower fitting algorithm with
pregenerated and Gaussian backgrounds. In order to perform the comparison the
configurations were optimized to obtain
fake rates at approximately the same value of 5\%, as shown
in \Cref{fig:fake-rate}. To obtain equal fake rates the shower fitting
algorithm parameter \texttt{TowerChi2ndfLimit} was set to 5.5 for pregenerated
background and 1.86 for Gaussian background, while the clusterization algorithm
parameters were set to their defaults.

A reconstructed cluster was considered to be a fake electron if it differs by
more than \SI{5}{\mrad} in $\theta$ and if the $\sin$ and $\cos$ of the azimuthal angle
differ by more than 0.35 with respect to the generated particle. The energy of
the reconstructed cluster was required to be above the configured threshold as
well. Because the fake rate only depends on the selection criteria and the
background, it is independent of the incident electron energy.

\begin{figure}
  \centering
  \begin{subfigure}[b]{0.48\textwidth}
    \includegraphics[width=\textwidth]{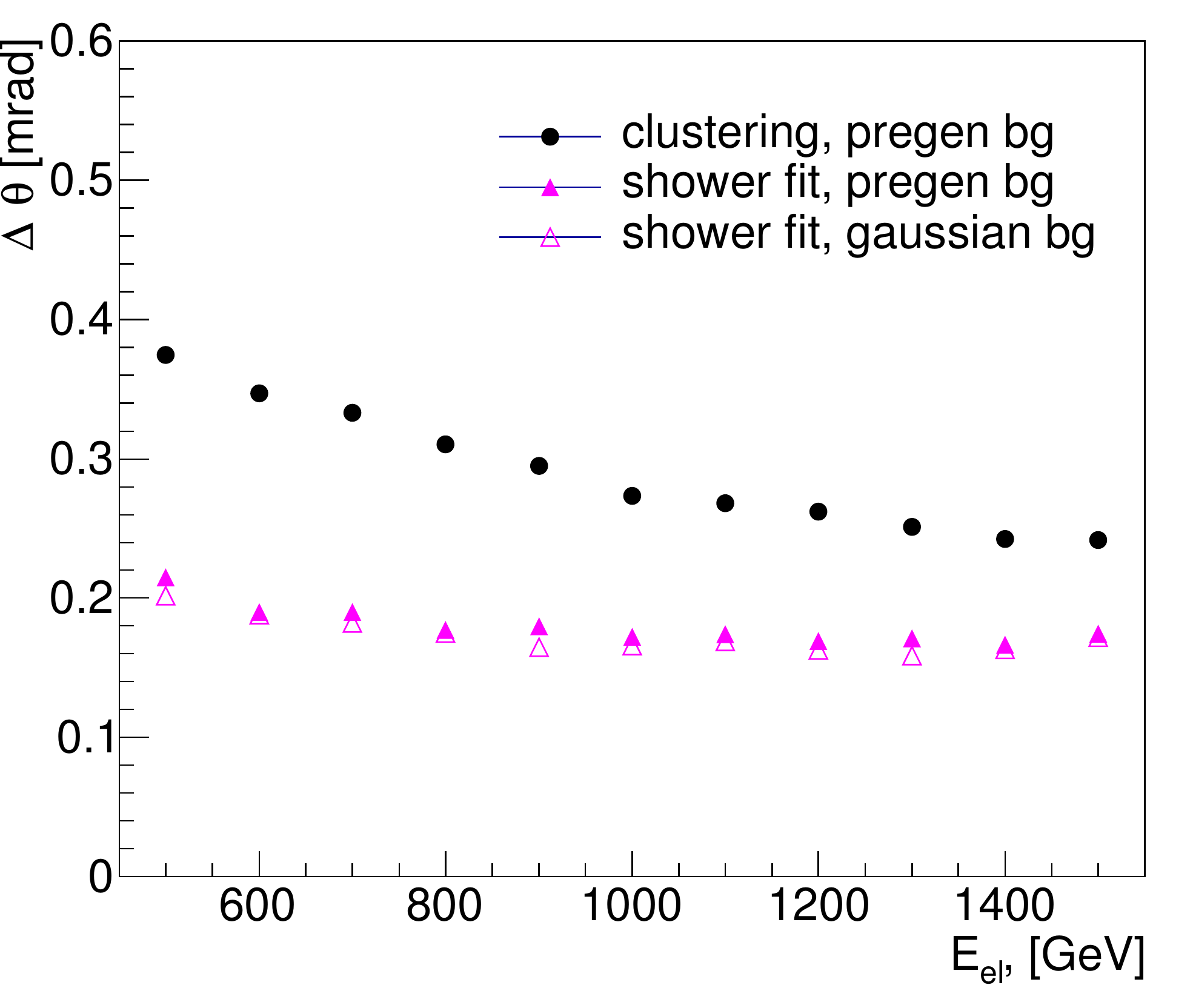}
    \subcaption{}\label{fig:theta-res}
  \end{subfigure}
  \hfill
  \begin{subfigure}[b]{0.48\textwidth}
    \includegraphics[width=\textwidth]{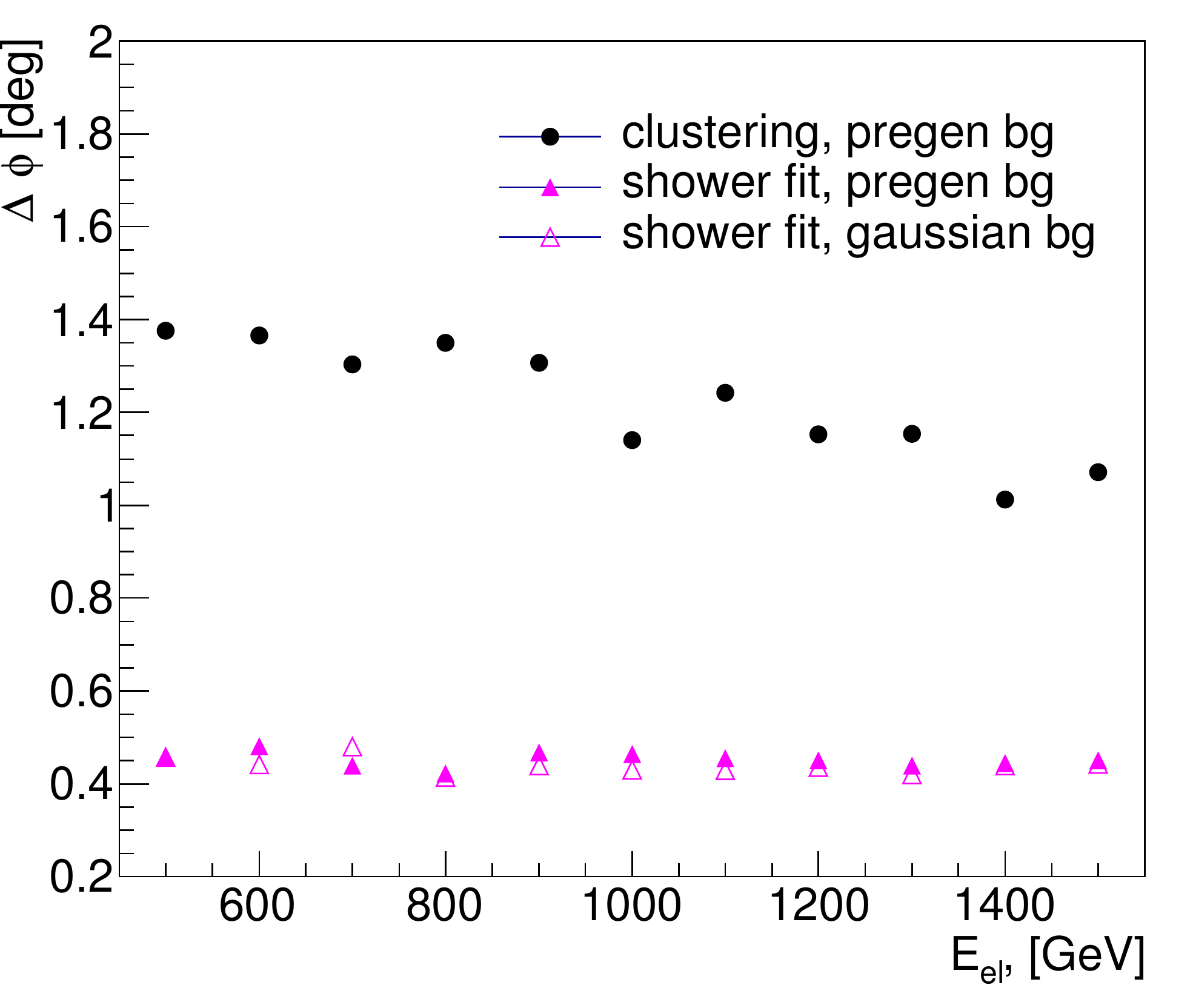}
    \subcaption{}\label{fig:phi-res}
  \end{subfigure}
  \caption{Angular resolution of reconstructed electrons for different
    reconstruction and background simulation methods as a function of the
    electron energy for the polar \subref{fig:theta-res} and azimuthal \subref{fig:phi-res}
    angle.\label{fig:spacial-res}}
\end{figure}

\Cref{fig:spacial-res} shows a resolution comparison for the two algorithms as
a function of the energy of the signal electron. The resolution is defined as
the standard deviation of the difference between the measured quantity and the
original value taken from the generator level. With the configuration
described above, the polar angle resolution is in the range of
\SIrange{0.25}{0.4}{\mrad} for the clustering-based algorithm and
\SIrange{0.2}{0.3}{\mrad} for the shower fitting approach. For the reconstruction
of the azimuthal angle, the clustering method gives an average of \ang{1.2} and the
shower
fitting method an average of \ang{0.5}. In both cases the resolution increases with
higher electron energy. As expected, the fitting algorithm being properly
adjusted shows better resolution than the clustering algorithm. The choice of the background
simulation has a negligible effect on the angular resolution.

\begin{figure}
  \centering
  \begin{subfigure}[b]{0.48\textwidth}
    \includegraphics[width=\textwidth]{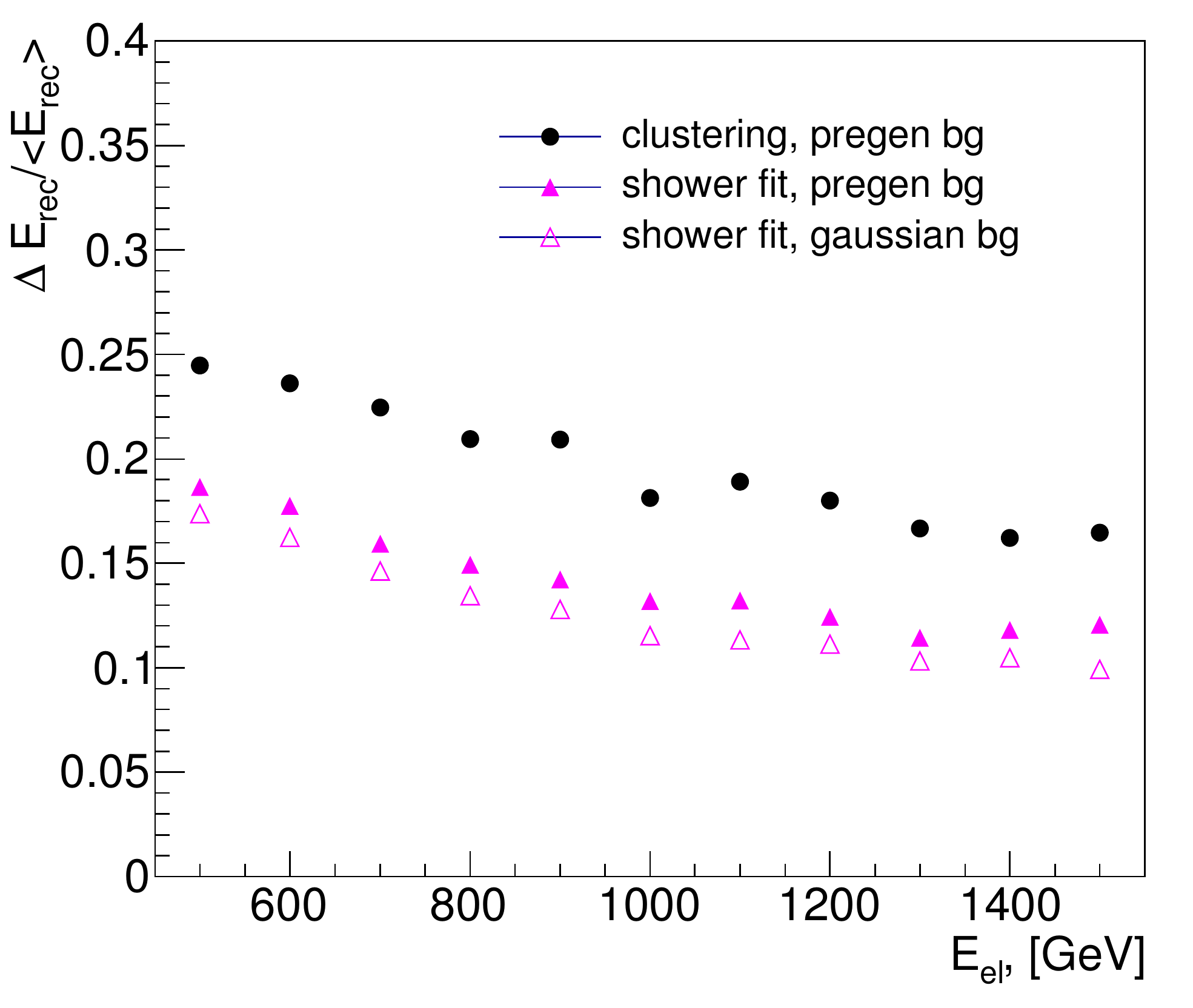}
  \end{subfigure}
  \caption{Energy resolution of reconstructed electrons for different
    reconstruction and background simulation methods as a function of the electron
    energy.}\label{fig:energy-res}
\end{figure}

The energy resolution is defined as the ratio of the standard deviation of the reconstructed
cluster or shower energy to its average. The resolution is shown in \cref{fig:energy-res}. The
plot shows that the shower fitting algorithm provides approximately 20\% better energy resolution
than the clustering algorithm. There is little sensitivity to the background generation
method. In all cases the resolution improves with higher electron energies.

\section{Summary}
\label{sec:summary}

A dedicated study of the reconstruction of high-energy electrons in the BeamCal
detector at CLIC and ILC is presented. The forward detector regions at these
future \Pep{}\Pem{} colliders will be exposed to high particle fluxes from beam-induced
background. This imposes constraints on the detection of signal particles at
small polar angles and requires optimised background simulation as well as
reconstruction algorithms.


To realistically approximate the reconstruction of high energy electrons in the
forward region a reconstruction package for the \marlin{} framework was
developed. It creates background distributions and reconstructs showers in the
BeamCal.

Four different methods to create background distributions were implemented and
compared: using pregenerated background distributions, parametrized
distributions, Gaussian approximation, and a method that is backward compatible
with the
existing BeamCal reconstruction in \textsc{MarlinReco}. While the methods vary
in complexity and performance, their impact on the reconstruction efficiency is
very small. Therefore each method can be used almost interchangeably. For precise
studies of the reconstruction efficiency the most realistic background creation
method is recommended. The method with the smallest resource requirements is
recommended for large scale Monte Carlo campaigns.

Two distinct algorithms for high-energy electron reconstruction in the BeamCal
are implemented in the reconstruction package and their performance was studied.
One algorithm is based on nearest neighbour clustering and the second one on the
shower shape fitting. A comparison of these algorithms show their consistency in
terms of reconstruction efficiency and resolution. They both can be configured
and optimised for the requirements of a specific physics analysis.

\newpage

\appendix
\section{Background Conversions}
\label{appendix:bkg-files}

To make use of the BeamCal reconstruction, a few steps are necessary to provide
the background in the appropriate format. First, the particles from the background
have to be simulated with a detector containing a BeamCal subdetector. A single
bunch crossing of background particles can be split into several individual
simulation events and files to speed up the simulation.

\subsection{Creating the Background Pool}
\label{sec:backpool}

To create the ROOT files from the simulated background particles, the
\texttt{ReadBeamCal} \marlin{} processor is part of the reconstruction package.
Each complete run of the processor will create a single ROOT file and all SLCIO
input files given to \marlin{} will be merged into a single bunch crossing.

The following parameters are used for the processor:

\begin{description}[topsep=8pt]
\item[BeamCalCollectionName =  BeamCalCollection  ]\  \\ name of the BeamCal Collection;
\item[OutputFileBackground = BeamCal.root ]\  \\ the name of the root file;
  containing the background bunch crossing;
\item[ProbabilityFactor = 100.0]\  \\ probability for a particle to be added
  to the bunch crossings. Allows the scaling of the background to a smaller
  background rate, for example to approximate the effect of beam--beam  
  offsets.
\end{description}

\subsection{Creating Background Parameter Files}
\label{sec:backpar}

In case of parametrized and Gaussian backgrounds the user has to supply a ROOT 
file with background parameters. The file is extracted from the background pool
with a tool which is a part of the reconstruction processor package. The compiled
package has an executable \texttt{BCBackgroundPar} in the \texttt{\$BCRECO/bin}
directory which should be run from the command line like:
\begin{verbatim}
  > $BCRECO/bin/BCBackgroundPar bckgrnd.root [[bckgrnd_2.root] ...]
\end{verbatim}
This command will produce a \texttt{BeamCal\_bg.root} file which has to be
specified in the processor configuration file (see \Cref{appendix:config}).

\newpage

\section{Reconstruction parameters}
\label{appendix:config}

The \marlin{} processor for the BeamCal reconstruction is named ``\texttt{BeamCalClusterReco}''. There are a
number of parameters which can be configured to reconstruct simulated events.
The given values are default for the package. 
\begin{description}[topsep=8pt]
\item[BeamCalCollectionName =  BeamCalCollection  ]\  \\ name of the BeamCal
  Collection;
\item[MCParticleCollectionName = MCParticle  ]\  \\ name of the Monte Carlo
  (generator-level) particles collection which is used to calculate total
  detector efficiencies;
\item[RecoClusterCollectionname = BCalClusters  ]\  \\ name of the
  Reconstructed Cluster collection;
\item[RecoParticleCollectionname = BCalRecoParticle  ]\  \\ name of the
  Reconstructed Particle collection; 
\item[CreateEfficiencyFile =  true   ]\  \\ flag to create reconstruction
  efficiency plots;
\item[EfficiencyFilename = TaggingEfficiency.root   ]\  \\ the name of the
  root-file which will contain the efficiency plots;
\item[BackgroundMethod = Gaussian  ]\  \\ defines background generation
  method. Possible values are: \texttt{Gaussian, Parametrised, Pregenerated,
    Averaged}. More details on the background definition are given in
    \Cref{sec:beam-bkg}.
\item[{InputFileBackgrounds =  [background\_file(s).root]}] \  \\ list 
  of the root-files with background information. In case of \texttt{Pregenerated}
  option selected, it specifies a list of background pool files each containing simulated
  background for single bunch crossing. In case of \texttt{Gaussian} or
  \texttt{Parametrised} background, it points to a single file with background
  parameters produced as described in \Cref{appendix:bkg-files}.
\item[MinimumTowerSize = 4   ]\  \\ in the clusterization approach this option
  defines a minimum number of pads for a single tower to be considered as part of
  a cluster;
\item[NumberOfBX = 40   ]\  \\ number of bunch crossings which fall into the
  read-out window. This value is used for background generation. For CLIC
  conditions the nominal value is 40, for ILC it is 1.
\item[PrintThisEvent = -1   ]\  \\ debug option to print event display for a
  given event number. The output is printed to eps-file in the current directory.
\item[UseConstPadCuts = false   ]\  \\ if \texttt{true}, the clusterization algorithm
  constructs clusters from pads satisfying cuts specified in the \texttt{ETPad}
  option. If \texttt{false}, the standard deviation of the background fluctuation
  in each pad is used multiplied by the \texttt{SigmaCut} factor.
\item[StartingRing = 0 1 2    ]\  \\ rings starting from which
  thresholds defined by \texttt{ETCluster} and \texttt{ETPad} are applied. I.e.,
  from ring 0 the first value is applied, from ring 1 the second from ring 2 the
  third. Can be an arbitrary number of values as long as \texttt{ETCluster} and
  \texttt{ETPad} contain the same number of values. Must start with 0.
\item[ETCluster = 3 2 1 ]\ \\ energy in a cluster/shower to consider it an
  electron (GeV). Each value corresponds to an entry in
  \texttt{StartingRing}. For the shower fitting approach or if the cut is proportional to standard background
  fluctuation (see \texttt{UseConstPadCuts} option above), then only the first value
  is used.
\item[ETPad = 0.5 0.3 0.2    ]\  \\ for clusterization approach, the values
  set lower limit on the pad energy after background subtraction. If
  \texttt{UseConstPadCuts} is \texttt{true} the first value is the minimum energy
  a pad has to contain to be considered. This option is not used in the shower fitting approach.
\item[SigmaCut = 3   ]\  \\ if \texttt{UseConstPadCuts} option is set to
  \texttt{false}, each pad with signal energy $E_{\tot} - E_{\bkg}$ above
  \texttt{SigmaCut}$\times\sigma_{\pad}$ is considered for clusters;
\item[StartLookingInLayer = 10   ]\  \\ layer starting from which the
  algorithms look for signal pads for both clusterization and shower fitting
  approach;
\item[NShowerCountingLayers = 3   ]\  \\ in the shower fitting approach, the
  pad energies are projected from layers between \texttt{Start\-Looking\-In\-Layer} and
  \texttt{NShowerCountingLayers}. See the algorithm description in \Cref{sec:shower-fit}.
\item[UseChi2Selection = true   ]\  \\ this option controls reconstruction
  algorithm: \texttt{false} for clusterization, \texttt{true} for shower
  fitting;
\item[TowerChi2ndfLimit =  2.0   ]\  \\ for the shower fitting algorithm, this is
  a limit on the square norm of projected pad energies
  $\chi^2/ndf$, where $\chi^2_{\proj} = \sum \frac{{(E_{\pad}^{\tot} -
    \bar{E}_{\pad}^{\bkg})}^2}{\sigma^2_{\pad}}$ and $ndf$ is a number of pads used
  for projection. Reasonable
  value for pregenerated background is 5, for Gaussian it is 2.
\end{description}

\newpage

\printbibliography[title=References]

\end{document}